\documentclass[epj]{svjour}
\usepackage[english]{babel}
\usepackage[latin1]{inputenc}
\usepackage[T1]{fontenc}
\usepackage{amssymb}
\usepackage{mathrsfs}
\usepackage[dvips]{graphicx}
\usepackage{epsfig,bm}

\newcommand{\beq}{\begin{equation}}
\newcommand{\eeq}{\end{equation}}

\begin{document}

\title{Stabilizing the intensity of a wave amplified by a beam of particles} 
\titlerunning{Stabilizing the intensity of a wave amplified by a beam of particles}
\date{\today}

\author{R.~Bachelard\inst{1}, A.~Antoniazzi\inst{2}, C.~Chandre\inst{1}, D.~Fanelli\inst{2,3}, X.~Leoncini\inst{1,4}, \and M.~Vittot\inst{1}
}
\authorrunning{R.~Bachelard et al.}
\offprints{R.~Bachelard}

\institute{Centre de Physique Th\'eorique\thanks{\emph{Present address:} Unit\'e Mixte de Recherche (UMR 6207) du CNRS, et des universit\'es Aix-Marseille I, Aix-Marseille II et du Sud Toulon-Var. Laboratoire affili\'e \`a la FRUMAM (FR 2291). Laboratoire de Recherche Conventionné du CEA (DSM-06-35).}, CNRS Luminy, Case 907, F-13288 Marseille Cedex 9, France 
\and Dipartimento di Energetica and CSDC, Università di Firenze, INFN, via S. Marta, 3, 50139 Firenze, Italy
\and Department of Cell and Molecular Biology, Karolinska Institute, SE-171 77 Stockholm, Sweden
\and Physique des Interactions Ioniques et Moléculaires, CNRS-Université de Provence, Centre de St Jérôme, F-13397 Marseille, France}

\date{Received: \today / Revised version: date}
\abstract{The intensity of an electromagnetic wave interacting self-consistently
with a beam of charged particles as in a free electron laser, displays large
oscillations due to an aggregate of particles, called the macro-particle. In
this article, we propose a strategy to stabilize the intensity by re-shaping
the macro-particle. This strategy involves the study of the linear stability
(using the residue method) of selected periodic orbits of a mean-field model. As
parameters of an additional perturbation are varied, bifurcations occur in the
system which have drastic effect on the modification of the self-consistent
dynamics, and in particular, of the macro-particle. We show how to obtain an
appropriate tuning of the parameters which is able to strongly decrease the
oscillations of the intensity without reducing its mean-value.}
\maketitle

\section{Introduction}\label{intro}

The self-consistent interaction between an electromagnetic wave (or a
set of electromagnetic waves) and a beam of charged particles is ubiquitous in
many branches of physics,
e.g. accelerator and plasma physics. For instance, it
plays a crucial role in the Free Electron Laser, which is
used to generate a tunable, coherent, high power radiations. Such
devices differ from conventional lasers in using a relativistic
electron beam as its lasing medium. The physical mechanism responsible
for the light emission and amplification is the
interaction between the beam and a wave, which
occurs in presence of a magnetostatic periodic field generated in an
undulator. Due to the effect of the magnetic field, the electrons are
forced to follow sinusoidal trajectories, thus emitting synchrotron
radiation. This initial seed, termed {\it spontaneous emission}, acts as a trap
for the electrons which in turn amplify it by emitting coherently, until the
laser effect is reached.

\par
The coupled evolution of radiation field and $N$ particles can
be modeled within the framework of a simplified Hamiltonian picture
\cite{bonifacio}. The $N+1$ degree of freedom Hamiltonian
displays a kinetic contribution, associated with the particles, and a
potential term accounting for the self-consistent coupling between the
particles and the field. Hence, direct inter-particle interactions
are neglected, even though an effective coupling is indirectly
provided because of the interaction with the wave.

The linear theory predicts \cite{bonifacio}, for the amplitude of the radiation
field, a
linear exponential instability and a late oscillating
saturation. Inspection of the asymptotic phase-space suggests that a
bunch of particles gets trapped in the resonance and forms a clump
that evolves as a single {\it macro-particle} localized in phase space. The
untrapped particles are almost uniformly distributed between two
oscillating boundaries, and populate the so-called {\it chaotic sea}.

Furthermore, the macro-particle rotates around a well defined center and this
peculiar dynamics is shown to be responsible for the
macroscopic oscillations observed for the intensity \cite{tennyson,antoniazzi}.
It can be
therefore hypothesized that a significant reduction in the intensity
fluctuations can be gained by implementing a dedicated control
strategy, aimed at confining the macro-particle in space. As a side
remark, note that  the size of the macro-particle is directly related
to the bunching parameter, a quantity of paramount importance in FEL
 context\cite{antoniazzi}.

For example, a static electric field \cite{tsunoda,morales,lin} can be
used to increase the average wave power. While the chaotic particles
are simply accelerated by the external field, the trapped ones are
responsible for the amplification of the radiation field.

The dynamics can also be investigated from a topological point of
view, by looking at the phase space structures. In the framework of a simplified
mean field description, i.e. the so-called {\it test-particle} picture where the particles passively
interact with a given electromagnetic wave, the trajectories of trapped
particles correspond to invariant tori, whereas
unbounded particles evolve in a chaotic region of phase-space. 
Thus, the macro-particle corresponds to a dense set of invariant tori. Our strategy is to
modify the macro-particle dynamics by restoring or destroying invariant tori in selected
regions of phase space.

A technique of Hamiltonian control can be used \cite{chandre,bachelard}
to reconstruct additional invariant tori around the macro-particle, in
order to enhance the trapping. A specific perturbation is computed,
which guarantees the confinement on invariant tori of trajectories
characterized by a specific energy.

In this paper, we propose a strategy to stabilize the intensity of the
wave, by modifying the characteristics of the macro-particle. A
(generic) one or several-parameter family of perturbations is introduced, which
allows us to
modify the topology of phase-space by tuning appropriately the parameters. The
residue method
\cite{greene,cary1,cary2,artres} is used to identify the important local
bifurcations happening in the system when the parameters are varied, by an analysis of
linear stability of selected periodic orbits. This technique
enables to monitor the size, gyration and internal structure of the
macro-particle. An appropriate tuning of the parameters is able to
strongly decrease the oscillations of the intensity without reducing its
mean-value.

The paper is organized as follows~: In Sec. \ref{test}, the
test-particle model is presented. The latter provides a simple
topological representation of the self-consistent interaction. In
Sec. \ref{rescrit}, the residue method is discussed: in particular we
show how local bifurcations can generate an enlargement  of the
macro-particle. In Sec. \ref{newsec}, we apply the above method to re-shape both
the size and the internal structure of the macro-particle. Finally, in sect \ref{concl} we draw our conclusions.

\section{Dynamics of a single particle}\label{test}

The dynamics of the wave-particle interaction has been described in
Ref.~\cite{bonifacio} by the $N$-body Hamiltonian, accounting for a 
kinetic contribution and an interaction term between the particles and the radiation field~:
\beq\label{HN}
H_{N}(\{\theta_j,p_j\},\phi,I) = \sum_{j = 1}^{N} \frac{p_j^{2}}{2} - 2 \sqrt{\frac{I}{N}} \sum_{j = 1}^{N} \cos{(\phi+\theta_j)},
\eeq
where $(\theta_j,p_j)$ are the conjugate phase and momentum of the $i$-th
particle, and $(\phi,I)$ stand respectively for the conjugate phase and
intensity of the radiation field. Since $\phi$ is a phase, $(\phi,I)$
belongs to ${\mathbb T}\times {\mathbb R}^{+}$ where $\mathbb T$ is
the one-dimensional torus. 
here $(\theta_j,p_j)$ belongs to ${\mathbb T}\times{\mathbb R}$. The phase
space of the system is then ${\mathbb T}^{N+1}\times {\mathbb R}^{N}\times
{\mathbb R}^{+}$. We notice that there are two conserved quantities~:
$H_N$ and the total momentum $P_N=I+\sum_j p_j$. We consider the
dynamics given by Hamiltonian (\ref{HN}) on a $2N$-dimensional
manifold (defined by $H_N=0$ and $P_N=\varepsilon$ where $\varepsilon$
is infinitesimally small).

\begin{figure}[t] 
  \centerline{
    \mbox{\includegraphics[width=3.2in, height=1.5in]{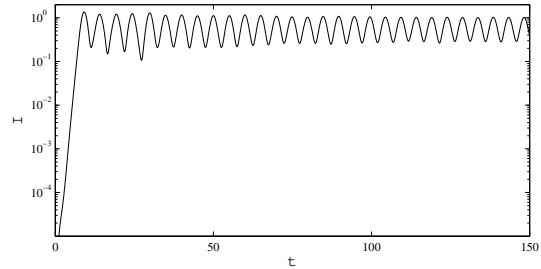}}}
  \caption{Normalized intensity $I/N$ from the dynamics of Hamiltonian (\ref{HN}), with $N=10000$ particles and $H_N=0$, $P_N=10^{-7}$.\label{int0}}
\end{figure}

\begin{figure}[t] 
  \centerline{
    \mbox{\includegraphics[width=3.2in, height=2.2in]{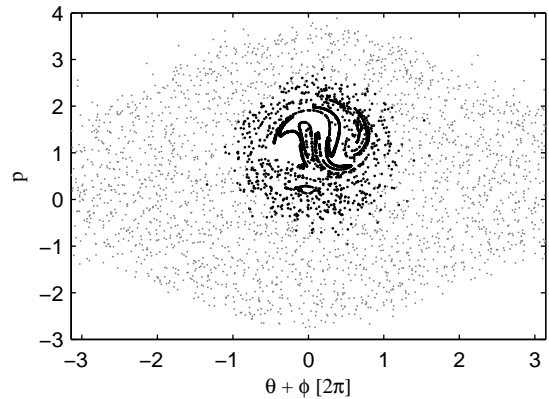}}}
  \caption{Snapshot of the $N$ particles at $t=1000$, with $N=10000$ (same parameters as Fig.~\ref{int0}). The grey points correspond to the chaotic particles, the dark ones to the particles in the macro-particle.\label{snap0}}
\end{figure}

Starting from a
negligible level ($I=\varepsilon$ small and $p_j=0$), the intensity grows exponentially and eventually
attains a saturated state characterized by large oscillations, as depicted in
Fig. \ref{int0}. Concerning the particle dynamics,
around half of them are trapped by the wave and form the so-called macro-particle (see Fig. \ref{snap0}). A particle is trapped if it travels in the potential well of the wave, after the system has reached saturation, which reads~:
\beq\label{trap}
\exists k \in \mathbb{Z} \mbox{such that} \forall t \geq t_{sat}, (2 k-1) \pi <\theta(t) + \phi(t)<(2 k+1)\pi,
\eeq 
The remaining particles experience a chaotic motion within an oscillating waterbag, termed chaotic sea, which is unbounded in $\theta$ contrary to the macro-particle.

\begin{figure}[t] 
  \centerline{
    \mbox{\includegraphics[width=3.2in,height=2in]{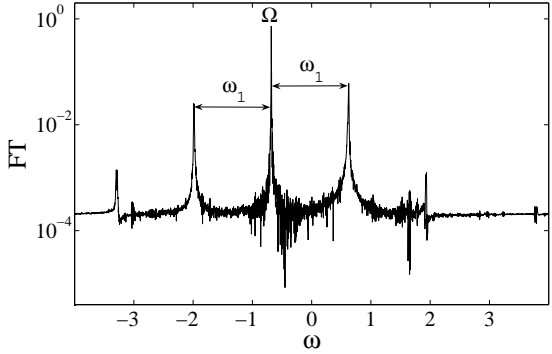}}}
  \caption{Fourier transform of the interaction term $h(t)$, as obtained from simulations of Hamiltonian (\ref{HN}), after saturation has been reached (same parameters in Fig. \ref{int0}).\label{ft}}
\end{figure}

In order to get a deeper insight into the dynamics, we consider the motion of a
single particle. For large $N$, we assume that its influence on the wave is negligible, thus it can be schematized as a passive 
particle in an oscillating field. The
motion of this test-particle is described by the following 
Hamiltonian with one and a half degrees of freedom~:

\begin{eqnarray}\label{H1P}
H_{1p} (\theta,p,t) & = & \frac{p^{2}}{2} - 2 \sqrt{\frac{I(t)}{N}} \cos{( \theta +\phi(t))} \\
& = & \frac{p^{2}}{2} - Re(h(t) e^{i \theta}),
\end{eqnarray}
where the interaction term $h(t)$ is derived from some simulations of the
original $N$-body Hamiltonian (\ref{HN}). In the saturated regime, $h(t)$ is mainly periodic (see Fig. \ref{ft}). A refined Fourier analysis \cite{laskar} shows that it can be written as~:
\begin{equation}\label{inter}
h(t) = 2 \sqrt{\frac{I(t)}{N}} e^{i \phi(t)} \approx [F + \alpha e^{i \omega_1 t} + \beta e^{-i \omega_1 t}] e^{i \Omega t},
\end{equation}
where $\Omega=-0.685$ stands for the wave velocity and $\omega_1=1.291$ for the
frequency of the oscillations of the intensity. As for the amplitudes, the Fourier analysis provides the following values~: $F=1.5382 - 0.0156i$, $\alpha=0.2696 - 0.0734i$ and $\beta=0.1206 + 0.0306i$.

Hamiltonian~(\ref{H1P}) results from a periodic perturbation of a pendulum described by the integrable Hamiltonian $H_0$
$$
H_0=\frac{p^2}{2}-\vert F\vert \cos(\theta+\Omega t+\phi_F),
$$
where $F=\vert F\vert e^{i\phi_F}$. The linear frequency of this pendulum is $\sqrt{\vert F\vert}\approx 1.240$ which is very close to the frequency of the forcing $\omega_1$. Therefore a chaotic behaviour is expected when the perturbation is added even with small values of the parameters $\alpha$ and $\beta$.

\begin{figure}[t] 
  \centerline{
    \mbox{\includegraphics[width=3.2in]{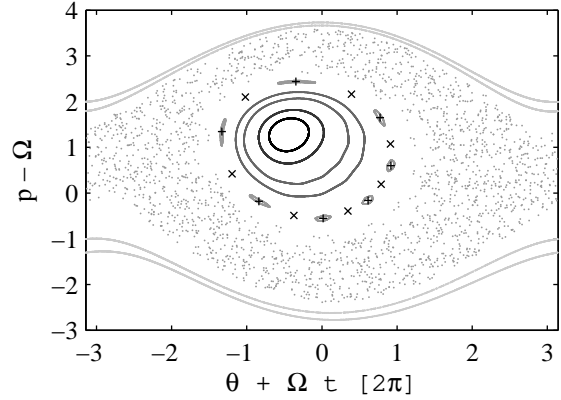}}}
  \caption{Poincaré section of a test-particle, described by Hamiltonian (\ref{H1P}). The periodic orbits with r.n. $7$ are marked by plus (elliptic orbit) and crosses (hyperbolic orbit).\label{QPc0}}
\end{figure}

The Poincaré sections (stroboscopic plot performed at frequency
$\omega_1$) of the test-particle (see Fig. \ref{QPc0}) reveals that the
macro-particle corresponds to a set of invariant
tori. Conversely, the chaotic sea is filled with seemingly erratic trajectories
of particles. The rotation of the macro-particle and the
oscillations of the waterbag can be visualized by translating
continuously in time the stroboscopic plot of the phase space. 

The macro-particle is organized around a central (elliptic) periodic orbit with rotation number (r.n.) $1$. The period of oscillations of the intensity is the same as the one of the macro-particle which indicates the role played by this coherent structure in the stabilization of the wave. 

Thus, in the test-particle model, the macro-particle is formed by particles which are trapped on two-dimensional invariant tori. This picture can be extended to the self-consistent model, if one considers the projection of a trajectory $(\phi(t),I(t),\{ \theta_j (t), p_j (t) \}_j)$ in the $(\theta,p)$ plane, each time it crosses the hyperplane $ \sum_{j} \sin{(\phi+\theta_j)} = 0$, i.e. $dI/dt=0$. From the full trajectory, we follow a given particle (an index $j$) and plot $(\theta_j,p_j)$ each time the full trajectory crosses the Poincar\'e section.

\begin{figure}[t] 
  \centerline{
    \mbox{\includegraphics[width=3.2in,height=2.2in]{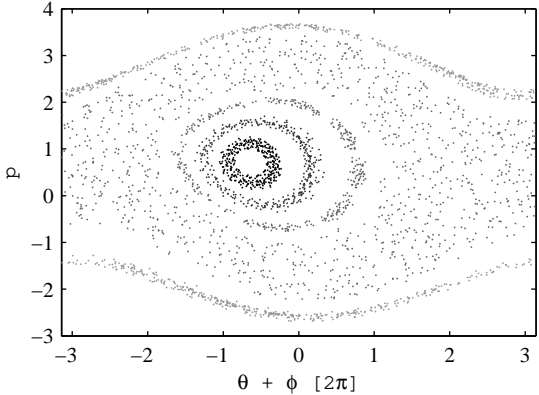}}}
  \caption{Poincar\'e section of Hamiltonian (\ref{HN}), when the particles intersect the plane $dI(t)/dt=0$. The different trajectories are represented by different grey levels.\label{itn}}
\end{figure}

The trapped particles (following definition (\ref{trap})) appear to be confined to domains of phase-space much smaller than the one of the macro-particle (see Fig. \ref{itn}). These domains are similar to the invariant tori of the test-particle model, although thicker.
It is worth noticing that not only these figures have a similar overall layout but there is a deeper correspondence in the structure of the macro-particle. For instance, both figures show a periodic orbit with period $7$ at the boundary of the regular region. Since we saw that the macro-particle directly influences the oscillations of the wave, the test particle Hamiltonian (\ref{H1P}) serves as a cornerstone of our control strategy which consists in modifying the structure of the macro-particle in order to stabilize the intensity of the wave. This strategy focuses on restoring or breaking-up invariant tori to reshape the macro-particle. In order to act on invariant tori, we use the periodic orbits which, as we have seen in Figs. 4 and 5 structure the macro-particle. 

\section{Residue method}\label{rescrit}

The topology of phase space is investigated by analysing the
linear stability of periodic orbits. Information on the nature of these orbits (elliptic,
hyperbolic or parabolic) is provided using, e.g., an indicator like Greene's residue
\cite{greene,mackay}, a quantity that enables to monitor local changes
of stability in a system subjected to an external perturbation 
\cite{cary1,cary2,artres,howard}.

Let us consider an autonomous Hamiltonian flow with two degrees of freedom
which depends on a set of parameters $\bm{\lambda}\in{\mathbb R}^m$~:

$$
\dot{\bm{z}}={\mathbb J}\nabla H(\bm{z};\bm{\lambda}),
$$
where $\bm{z}=(p,E,\theta,t)\in{\mathbb R}^4$ and ${\mathbb J}=\left(
\begin{array}{cc} 0 & -{\mathbb I}_2 \\ {\mathbb I}_2 & 0
\end{array}\right)$, and ${\mathbb I}_2$ being the two-dimensional
identity matrix. In order to analyse the linear stability properties
of the associated periodic orbits, we also consider the tangent flow written as

$$
\frac{d}{dt}{J^t}(\bm{z})={\mathbb J}\nabla^2 H(\bm{z};\bm{\lambda}) J^t,
$$ 
where $J^0={\mathbb I}_4$ and $\nabla^2 H$ is the Hessian matrix
(composed of second derivatives of $H$ with respect to its canonical
variables). For a given periodic orbit with period $T$, the linear
stability properties are given by the spectrum of the monodromy matrix
$J^T$. These properties can be synthetically captured
in the definition of Greene's residue~:

$$ R=\frac{4-\mbox{tr}J^T}{4}.$$
In particular, if $R\in ]0,1[$, the periodic orbit is called elliptic (and is in general stable); if
$R<0$ or $R>1$ it is hyperbolic; and if $R=0$ and $R=1$, it is
parabolic while higher order expansions give the stability of such periodic orbits.

 Since the periodic orbit and its stability depend on the set of
 parameters $\bm{\lambda}$, the features of the dynamics will change
 under apposite variations of such parameters. Generically, periodic orbits and their (linear or nonlinear) stability properties are
 robust to small changes of parameters, except at specific values when
 bifurcations occur. The residue method \cite{cary1,cary2,artres} detects the
 rare events where the linear stability of a given periodic orbit changes thus allowing to calculate the appropriate values of
the parameters leading to the prescribed behaviour of the dynamics. As a consequence, this method can 
 yield reduction as well as enhancement of chaos.

 First, we illustrate the method by introducing an additional
 interaction term which depends on a parameter. The latter has to be properly tuned in order to control the
 dynamics. Here, the {\it control term} is chosen as
\beq\label{h1pc1n2}
H_{1p}^{c} (\theta,p,t;\lambda) = H_{1p} (\theta,p,t) - 2 \lambda \sqrt{\frac{I(t)}{N}} \cos{(2\theta +\phi(t))},
\eeq
and it is therefore similar to the original
 interaction term between the charged particle and the wave.

Alternatively, other types of perturbations could be selected, our choice being solely motivated by didactic reasons. For $\lambda=0$ (which corresponds to the original Hamiltonian $H_{1p}$), we consider two Birkhoff
 periodic orbits which originates from the breakup of invariant tori with rational ratio of the integrable case (given by the nonlinear pendulum described by $H_0$ in the previous section). These two periodic orbits have the same action but different
 angles in the integrable case $H_0$ and have the same r.n.
 on the Poincar\'e section; one is elliptic ${\mathcal
 O}_e$ and one is hyperbolic ${\mathcal O}_h$ (see Fig.~\ref{QPc0}). Let us recall that the rotation number (or winding number) of a periodic orbit is the number of times it crosses the Poincar\'e section before closing back onto itself.
 
\begin{figure}[t] 
  \centerline{
    \mbox{\includegraphics[width=3.2in,height=1.9in]{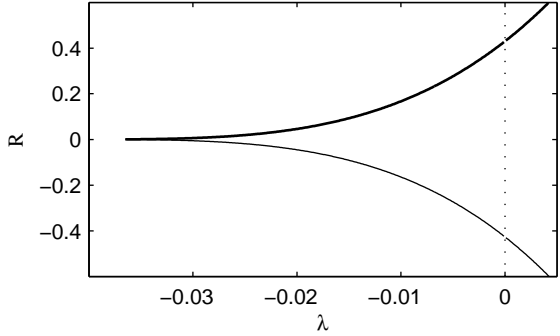}}}
  \caption{Residue of the elliptic (bold curve) and hyperbolic points with r.n. $7$ of Hamiltonian~(\ref{H1P}).\label{resc0n2}}
\end{figure}

\begin{figure}[t] 
  \centerline{
    \mbox{\includegraphics[width=3.2in,height=2.2in]{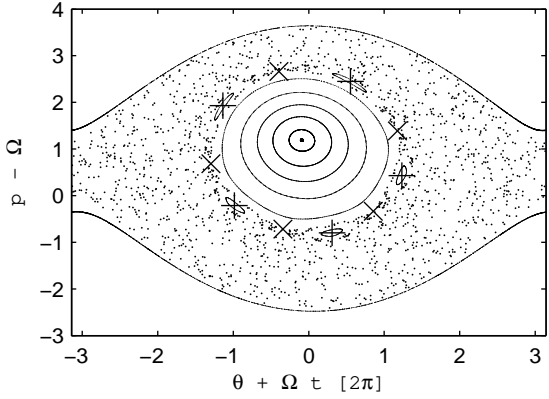}}}
  \caption{Poincaré section of Hamiltonian~(\ref{h1pc1n2}) with $\lambda=\lambda_c \approx -0.0370$. The periodic orbits with r.n. $5$ are marked by plus (elliptic orbit) and crosses (hyperbolic orbit).\label{qpc1n2}}
\end{figure}

\begin{figure}[t] 
  \centerline{
    \mbox{\includegraphics[width=3.2in,height=2.2in]{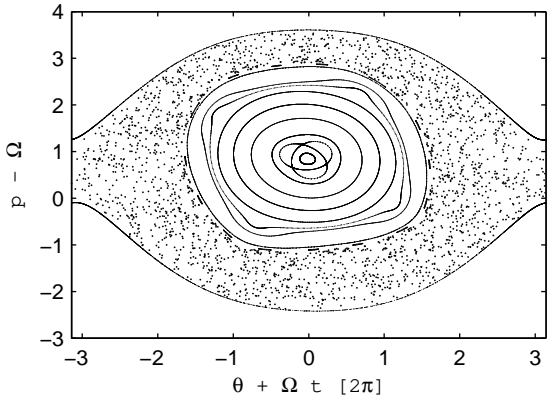}}}
  \caption{Poincaré section of Hamiltonian (\ref{h1pc1n2}) with $\lambda=\lambda_c^{\prime\prime} \approx -0.1321$.\label{qpc3n2}}
\end{figure}

  We call $R_e$ and $R_h$ the residues of these orbits. We have $R_e(0)>0$ and
 $R_h(0)<0$. We then modify continuously the parameter $\lambda$ starting from $0$. For each value of $\lambda$, we follow continously the periodic orbit under consideration as well as its linear stability property indicated by its residue. We plot the values of the residues as a function of the parameter $\lambda$ in
 Fig.~\ref{resc0n2}. We notice that at $\lambda=\lambda_c\approx -0.0370$ we have~:
\beq\label{bif00}
R_e(\lambda_c)=R_h(\lambda_c)=0.
\eeq
The bifurcation (\ref{bif00}) is associated with
 the creation of an invariant torus \cite{artres}. This diagnostic is
 confirmed by the Poincaré section (see Fig. \ref{qpc1n2}) of the {\it
 controlled} Hamiltonian (\ref{h1pc1n2}), at $\lambda=\lambda_c$~: The
 elliptic islands with r.n. $7$ have been replaced by a set of
 invariant tori in its neighborhood, leading to an enlargement of the macro-particle. Note
 that elliptic islands with r.n. $5$ are now present around the
 regular core (these orbits were present in the case $\lambda=0$, but were both hyperbolic in the chaotic sea). 
The associated couple of elliptic/hyperbolic orbits can be treated
similarly as those of r.n. $7$ in order to gain further enlargement
of the macro-particle~: we modify the value of the parameter $\lambda$ of Hamiltonian (\ref{h1pc1n2})
around $\lambda=\lambda_c$, until the condition
(\ref{bif00}) holds for
the residues of the considered orbits. Such a condition is verified
when $\lambda =\lambda_c^\prime\approx-0.0746$, and leads to the apparition of a new set
of invariant tori (not shown here); again it appears that elliptic islands with r.n. $4$
surround the macro-particle. The process of increasing the size of the macro-particle can be iterated one step further. 
We modify the parameter $\lambda$ around $\lambda=\lambda_c^{\prime}$ such that condition (\ref{bif00}) holds for the periodic orbits with r.n. $4$. The residue method predicts the formation
of an invariant torus at $\lambda_c^{\prime
  \prime}\approx -0.1321$. Inspection of phase-space (see
Fig.~\ref{qpc3n2}) corroborates our findings, resulting in an
additional extension in size of the macro-particle.

\begin{figure}[t] 
  \centerline{
    \mbox{\includegraphics[width=3.2in,height=2.2in]{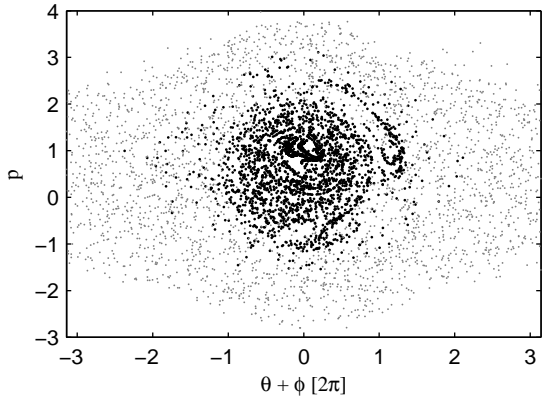}}}
  \caption{Snapshot of the dynamics in the $(\theta,p)$-plane at
  $t=1000$, when Hamiltonian (\ref{hn2}) is simulated with $N=10000$
  particles and $\lambda=-0.1321$. Here the control is switched on at  
  $t=300$, i.e. after saturation has been achieved. The grey points correspond to the
  particles of the chaotic sea, the dark ones form the macro-particle
  [see Eq.~(\ref{trap})].\label{snapcn2}}

\end{figure}

The test-particle approach is validated using the full self-consistent Hamiltonian involving $N$
particles (with $N \gg 1$) interacting with the wave. The control is
naturally introduced in the self-consistent dynamics as~:

\begin{eqnarray}\label{hn2}
H_{N}^{c}(\{\theta_j,p_j\},\phi,I;\lambda) & = & H_{N}(\{\theta_j,p_j\},\phi,I) \nonumber \\
& & - 2 \lambda \sqrt{\frac{I}{N}} \sum_j  \cos{( 2\theta_j +\phi)},
\end{eqnarray}
where $H_N$ is given by Eq.~(\ref{HN}).
The behaviour of the system is investigated for the three critical values where bifurcations happen in the test-particle model $\lambda_c$, $\lambda_c^\prime$, and $\lambda_c^{\prime
  \prime}$. For all these cases, simulations based on Eq.~(\ref{hn2})
display a qualitative behaviour similar to the one obtainde for the original unperturbed dynamics~(\ref{HN}): first a transient regime is detected where the intensity grows exponentially, followed by a subsequent saturation. This regime is characterized by the formation of a macro-particle whose dynamics is responsible for the oscillations
observed at the intensity level. 
Importantly, the macro-particle is shown to increase in size also when
operating within the framework of the relevant self-consistent context
(see Fig.~\ref{snapcn2}). In order to quantitatively characterize the evolution, we
define a radius and a gyroradius associated with the inner massive aggregate. The
radius corresponds to the standard deviation of the trapped particles, namely~:

\beq
R(t)=\sqrt{\langle({\bf r}_j(t) - {\bf r}_G(t))^2\rangle_{j}},
\eeq
where ${\bf r}_j(t)=(\theta_j(t), p_j(t))$ stands for the coordinates
of the $j$-th particle in phase-space, ${\bf r}_G(t)=\langle{\bf
  r}_j(t)\rangle_{j}$ represents the center of mass of the
macro-particle, and $\langle\rangle_{j\in \mathcal{M}}$ denotes the
average over the subset of particles constituting the inner core. 
On the other hand, the gyroradius characterizes the
rotation of the macro-particle and reads~:

\beq
R_{gyr}=\sqrt{\langle({\bf r}_G(t)-\overline{{\bf r}_G})^2\rangle_{t \geq t_{sat}}},
\eeq
where $\langle \rangle_{t \geq t_{sat}}$ denotes an averaging over time (after saturation has been reached at $t_{sat}$), and
$\overline{{\bf r}_G}=\langle{\bf r}_G(t)\rangle_{t \geq t_{sat}}$. As it is shown in
Table \ref{tabc2} the radius $R$
which is almost constant in time, increases by $28 \%
$ when the control is
turned on at $\lambda =
\lambda_c^{\prime \prime} $, whereas the gyroradius is significantly
reduced (by nearly a factor $6$). These indicators points to an effective 
stabilization of the macro-particle dynamics in phase-space.

\begin{figure}[t] 
  \centerline{
    \mbox{\includegraphics[width=3.2in,height=1.6in]{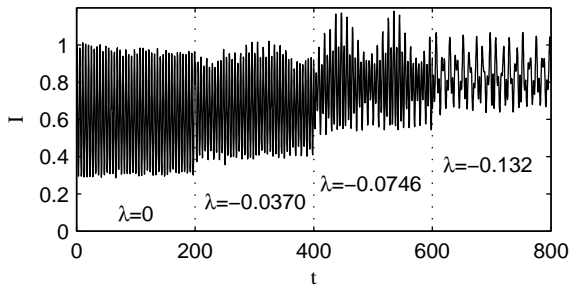}}}
  \caption{Intensity of the wave in the saturated regimes for Hamiltonian~(\ref{hn2}) for four values of the parameter~: $\lambda = 0$, $-0.037$, $-0.0746$ and $-0.1321$. For each value of $\lambda$, the system starts with a negligible value for the intensity, and a monokinetic beam of particles.\label{intc2}}
\end{figure}

As reference to the wave, the control induces an effective
stabilization of the 
intensity (see Fig. \ref{intc2}). In order to quantify the beneficial effects of
the control we measure two quantities, namely the mean value of the
intensity $\langle I \rangle = \langle I(t)
\rangle_{t \geq t_{sat}}$, and the average fluctuations $\Delta I = \langle
|I(t)-\langle I \rangle| \rangle_{t \geq t_{sat}}$. As confirmed by inspection of
Table \ref{tabc2} the value $\langle I
\rangle$ is increased by $34\%
$ when the control is turned on, while the
oscillations have damped by $39\%
$.

\begin{table}[t]
\centering
	\begin{tabular}{|c|c|c|c|c|}
	\hline
	$\lambda$ & 0 & -0.037 & -0.0746 & -0.1321 \\ \hline
	$R$ & 0.7 & 0.7 & 0.8 & 0.9 \\ \hline
	$R_{gyr}$ & 0.17 & 0.15 & 0.11 & 0.03 \\ \hline
	$\langle I \rangle$ & 0.62 & 0.66 & 0.79 & 0.83 \\ \hline
	$\Delta I$ & 0.73 & 0.66 & 0.66 & 0.44 \\ \hline
	\end{tabular}\caption{Characteristics of the macro-particle and the wave for Hamiltonian (\ref{hn2}).\label{tabc2}}
\end{table}

\section{Towards an effective stabilization of the intensity}\label{newsec}

In the previous section we provided numerical evidence that our method
enables to modify the size and position of the macro-particle. Here we
shall take one step further and demonstrate that we can 
alter the internal structure of the massive agglomerate. 
In particular, we will show that the macro-particle, originally composed of
invariant tori, can be chaotized, thus inducing an effective mixing while keeping the particles trapped.

Information on the internal structure of the macro-particle can be
gained from the residue of its central elliptic point, in the
test-particle picture. In particular, the destruction of invariant
tori can be expected if this orbit turns hyperbolic \cite{artres}. Thus,
the residues allows one to trace the following two prescribed behaviours
for the system~:
\begin{enumerate}
\item the chaotization of the macro-particle is controlled by the
  residue of the central periodic point (with r.n. $1$),
\item the trapping of particles is still guaranteed
  by the existence of invariant tori at the borders of the
  macro-particle (see Sec.~\ref{rescrit}).
\end{enumerate}
To this aim, the residue method is implemented with a control term containing additional parameters. We consider the following family of Hamiltonians~:
\beq\label{h1pck}
\tilde{H}_{1p}^{c}(\theta,p,t;{\bm \lambda}) =H_{1p}(\theta,p,t) - 2 \sqrt{\frac{I(t)}{N}} \sum_{k=2}^K \lambda_k \cos{(k\theta+\phi(t))},
\eeq
with ${\bm \lambda}=(\lambda_2,\lambda_3 \ldots \lambda_K)$, and where $H_{1p}$ is given by Eq.~(\ref{H1P}). Hence,
the original system is recovered for ${\bm \lambda}={\bf
  0}$. 
    
\begin{figure}[t] 
  \centerline{
    \mbox{\includegraphics[width=3.2in,height=1.9in]{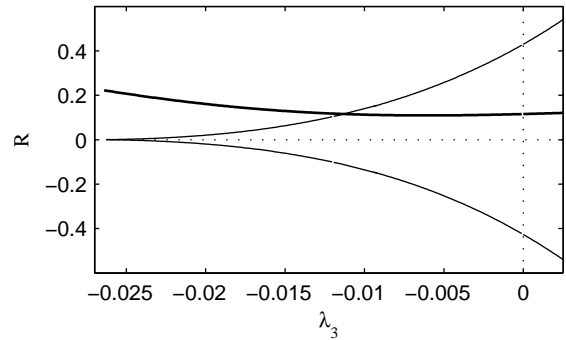}}}
  \caption{Residues of the elliptic (upper curve) and hyperbolic (lower curve) points with r.n. $7$, and residue of the periodic point with r.n.~$1$ (bold curve) of Hamiltonian~(\ref{h1pck}) With all the parameters set to zero except $\lambda_3$.\label{resc1n3}}
\end{figure}

First, the macro-particle is enlarged, while keeping its
core composed of invariant tori~: the system is perturbed by
tuning the parameter $\lambda_3$ around $0$; the residues of the
elliptic and hyperbolic periodic orbits with r.n. $7$ are monitored until condition
(\ref{bif00}) is reached at $\lambda_3^c\approx -0.0263$ (see
Fig.~\ref{resc1n3}), which in turn corresponds to restoring the invariant tori. Meanwhile, the residue of the central elliptic periodic point (with
r.n. $1$) remains stable.

The Poincar\'e section of Hamiltonian (\ref{h1pck}) with $\lambda_3=\lambda_3^{c}$ and $\lambda_{k\not= 3}=0$,
corroborates these predictions (not shown here, but similar to
Fig.~\ref{qpc1n2}, though the external elliptic islands have r.n. $6$
instead of $5$).

In order to get the macro-particle chaotic, we perturb the system
(by making $\lambda_6$ different from $0$) until the residue of the central
elliptic orbit (of r.n. $1$) crosses $0$ or $1$, whereas the residues
of the periodic orbits of r.n. $6$ satisfies
condition(\ref{bif00}). 

\begin{figure}[t] 
  \centerline{
    \mbox{\includegraphics[width=3.2in,height=1.9in]{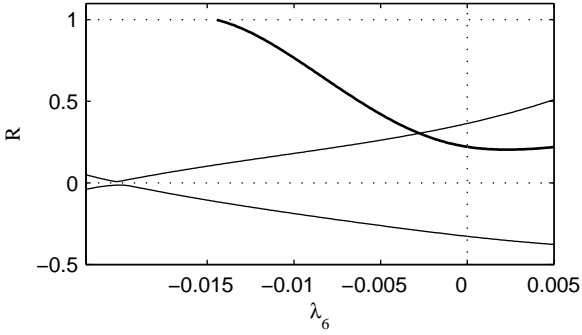}}}
  \caption{Residues of the elliptic (upper curve) and hyperbolic
  (lower curve) points with r.n. $6$, and residue of the central
  periodic point with r.n. $1$ (bold curve) for Hamiltonian~(\ref{h1pck}) with $\lambda_3=\lambda_3^c$ and $\lambda_{k\not= 3,6}=0$ as functions of $\lambda_6$. \label{3resn6}}
\end{figure}

Though the latter condition cannot be exactly matched for small values
of $\lambda_6$, the residue $R$ is shown to attain its minimal value
at $\lambda_6^c = -0.0201$ (see Fig. \ref{3resn6}), and one can 
expect a regularization of the dynamics in this region of
phase-space. Meanwhile, the residue of the central elliptic orbit has
crossed $1$, so one might expect chaos in this domain. Notice that
it was not possible to track numerically this central periodic orbit
beyond $\lambda_6^c$. Different scenarios are compatible with this finding. Possibly the latter has been broken
or alternatively it is associated with a tiny basin of attraction which prevents the multi-shooting Newton-Raphson to converge.

\begin{figure}[t] 
  \centerline{
    \mbox{\includegraphics[width=3.2in,height=2.2in]{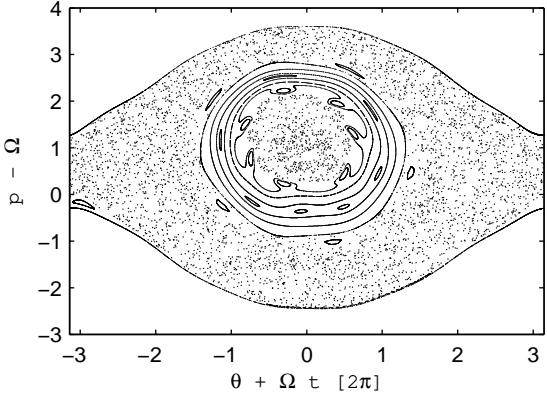}}}
  \caption{Poincaré section of Hamiltonian (\ref{h1pck}) with $\lambda_3=-0.0263$, $\lambda_6=-0.0201$ and $\lambda_{k\not= 3,6}=0$.\label{qpc2n6}}
\end{figure}

The Poincar\'e section for Hamiltonian~(\ref{h1pck}) with $\lambda_3=\lambda_3^c$, $\lambda_6=\lambda_6^c$ and $\lambda_{k\not= 3,6}=0$, depicted in Fig.~\ref{qpc2n6} displays a chaotic
core, while the invariant tori of its borders are preserved, ensuring
that the particles remain trapped. We have computed the Lyapunov exponents in
the different regions of phase space~:
   Inside the macro-particle it is equal to $0.06$ whereas, in the chaotic sea it is $0.14$, i.e., more unstable.

\begin{figure}[t] 
  \centerline{
    \mbox{\includegraphics[width=3.2in,height=2.2in]{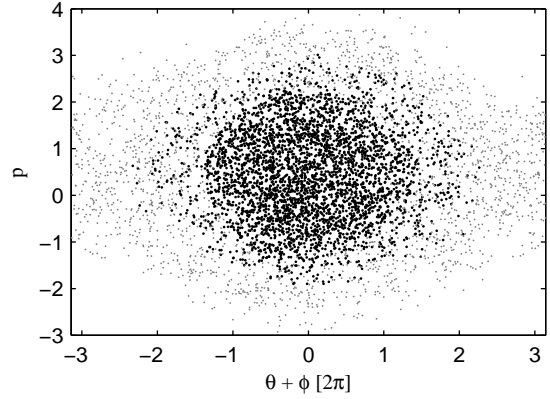}}}
  \caption{Snapshot of the dynamics in the $(\theta,p)$-plane of the
  $N=10000$ particles at $t=1000$ (the control started at $t=300$) for
  Hamiltonian~(\ref{hnck}) with $\lambda_3 = -0.0263$ and $\lambda_6 =
  -0.0201$. The control term is applied at $t=300$, after saturation has occured. The grey points correspond to the particles of the chaotic sea, the dark ones form the macro-particle.\label{snapcnk}}
\end{figure}

In order to complete the analysis, a set of
  simulations are performed within the self-consistent framework where $N$ particles interact with the wave. The additional perturbation is
naturally introduced as~:

\begin{eqnarray}\label{hnck}
\tilde{H}_{N}^{c}(\{\theta_j,p_j\},\phi,I;{\bm \lambda}) & = & H_{N}(\{\theta_j,p_j\},\phi,I) \nonumber \\
& & - 2 \sqrt{\frac{I}{N}} \sum_{k,j} \lambda_k \cos{(k\theta_j+\phi)},
\end{eqnarray}
where $H_N$ is given by Eq.~(\ref{HN}). Simulations in the two regimes of control, $(\lambda_3=\lambda_3^c, \lambda_{k \neq 3}=0)$ and $(\lambda_3=\lambda_3^c, \lambda_6=\lambda_6^c, \lambda_{k \neq 3,6}=0)$, confirm the prediction of the mean-field framework~: the macro-particle has been increased and homogenized (see Fig.~\ref{snapcnk}).

\begin{figure}[t] 
  \centerline{
    \mbox{\includegraphics[width=3.2in,height=1.6in]{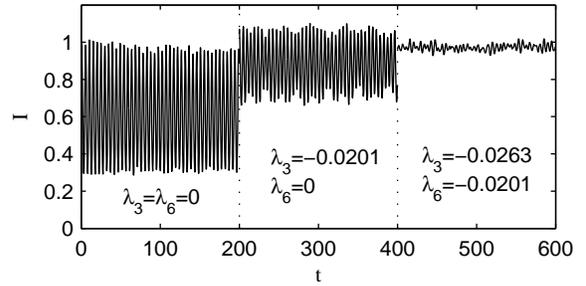}}}
  \caption{Intensity of the wave in the saturated regime for Hamiltonian (\ref{hnck}) with $\lambda_{k\not= 3,6}=0$. For each value of ${\bm \lambda}$, the system starts with a negligible value for the intensity, and a monokinetic beam of particles.\label{intck}}
\end{figure}

According to our estimates, the radius has been increased by more than $50\%
$, while its gyroradius has been decreased by two orders of magnitude (see Tab.~\ref{tabck}). As for the intensity of the wave (see Fig.~\ref{intck}), the fluctuations have been reduced by a factor $40$, while the mean-value has been raised by $50\%
$.
\begin{table}[t]
\centering
	\begin{tabular}{|c|c|c|c|}
	\hline
$\lambda_3$ & 0 & -0.0263 & -0.0263 \\ \hline
$\lambda_6$ & 0 & 0 & -0.0201 \\ \hline
$R$ & 0.7 & 1.15 & 1.22 \\ \hline
$R_{gyr}$ & 0.17 & 0.03 & 0.0015 \\ \hline
$\langle I \rangle$ & 0.62 & 0.88 & 0.97 \\ \hline
$\Delta I$ & 0.73 & 0.44 & 0.017 \\ \hline	
	\end{tabular}\caption{Characteristics of the macro-particle and the wave for Hamiltonian~(\ref{hnck}) with $\lambda_{k\not= 3,6}=0$.\label{tabck}}
\end{table}

\begin{figure}[t] 
  \centerline{
    \mbox{\includegraphics[width=3.2in,height=2.2in]{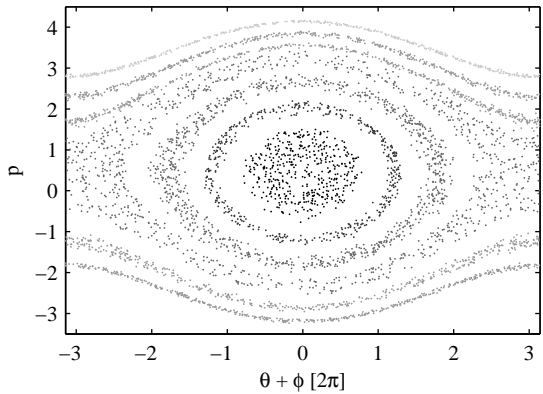}}}
  \caption{Poincar\'e section of Hamiltonian (\ref{hnck}) with $\lambda_3 = -0.0263$ and $\lambda_6 = -0.0201$, when the trajectory intersects the plane $dI(t)/dt=0$. The different trajectories are represented by different grey levels.\label{itnck}}
\end{figure}

In terms of topology of phase-space, it is interesting to observe that
the regularization of the dynamics is also found within the framework
of the original self-consistent picture (see Fig.~\ref{itnck}). Indeed, the confinement of the
trapped particles for Hamiltonian (\ref{h1pck}) to small domains of
phase-space is more pronounced than in the unperturbed case of
Hamiltonian (\ref{HN}).

\section{Conclusion}
\label{concl}

In this paper, we proposed a stabilization strategy in a wave-particle system and
 in particular in the a FEL setting. The technique uses an appropriate
 perturbation of the associated microscopic dynamics. We carried out
 our approach in the saturated regime of this system where the
 macro-particle, resulting from a set of invariant tori, plays a
 crucial role. In the framework of a mean-field approximation, a linear stability analysis of selected periodic orbits (coined by a residue method) can
point at the creation and destruction of invariant tori, allowing to re-shape the
macro-particle through specific values of the parameters of the
 additional perturbation. Numerical simulations performed for the
 original $N$-body 
self-consistent picture confirm the validity of the proposed approach 
and clearly demonstrate that the method allows one to predict a set of parameters for which a significant 
reduction of the oscillations of the intensity is found. Note that the residue method is
quite flexible and applies to any two degree of freedom Hamiltonian
 systems. We expect that other families of perturbation terms would yield similar results. 

As a final remark, we observed that the chaotization of the center of
the macro-particle (by a bifurcation of the central elliptic periodic
orbit which organizes the macro-particle) is associated with a strong
stabilization of the intensity.

\section*{Acknowledgements}

This work is supported by Euratom/CEA (contract EUR~344-88-1~FUA~F). 
We acknowledge useful discussions, comments and remarks 
from G.~De~Ninno, 
Y.~Elskens and the Nonlinear Dynamics team at CPT.

\end{document}